\documentclass[a4paper]{jpconf}
\usepackage{graphicx}
\usepackage{amsmath, amssymb, setspace, mathrsfs}

\newcommand\ket[1]{\left| #1 \right>}
\newcommand\braket[2]{\left< #1 | #2 \right>}
\newcommand\abs[1]{\left| #1 \right|}

\def\dd{\mathrm{d}}

\def\dsH{\mathrm{I}\hspace{-2.25pt}\mathrm{H}}
\def\dsM{\mathrm{I}\hspace{-2.25pt}\mathrm{M}}
\def\dsGamma{\mathrm{I}\hspace{-2.25pt}\Gamma}

\begin{document}
\title{Stimulated X-rays in resonant atom Majorana mixing}

\author{A Segarra and J Bernab\'eu}

\address{Departament de F\'isica Te\`orica \& IFIC, Universitat de Val\`encia - CSIC, C/ Dr. Moliner 50, E-46100 Burjassot (Spain)}

\ead{alejandro.segarra@uv.es, jose.bernabeu@uv.es}

\begin{abstract}
	Massive neutrinos demand to ask whether they are Dirac or Majorana
	particles.
	Majorana neutrinos are an irrefutable proof of physics beyond the Standard
	Model.
	Neutrinoless double electron capture is not a process but a virtual $\Delta
	L=2$ mixing
	between a parent $^AZ$ atom and a daughter $^A(Z-2)$ excited atom with two
	electron holes.
	As a mixing between two neutral atoms and the observable signal in terms of
	emitted two-hole X-rays,
	the strategy, experimental signature and background are different from
	neutrinoless double beta decay.
	The mixing is resonantly enhanced for almost degeneracy and, under these
	conditions,
	there is no irreducible background from the standard two-neutrino channel.
	We reconstruct the natural time history of a nominally stable parent atom
	since its production either by nature or in the laboratory.
	After the time periods of atom oscillations and the decay of the
	short-lived daughter atom,
	at observable times the relevant ``stationary" states are the mixed
	metastable long-lived state
	and the short-lived excited state, as well as the ground state of the
	daughter atom.
	Their natural population inversion is most appropriate for exploiting the
	bosonic nature
	of the observed X-rays by means of stimulating X-ray beams.
	Among different observables of the atom Majorana mixing,
	we include the enhanced rate of stimulated X-ray emission from the
	long-lived
	metastable state by a high-intensity X-ray beam.
	A gain factor of 100 can be envisaged in a facility like European XFEL.

\end{abstract}

\section{Introduction}
The experimental evidence of neutrino oscillations \cite{1,2} implies that
neutrinos are massive particles and that the three flavor neutrinos are
mixtures of the neutrinos with definite masses. The existence of non-vanishing
neutrino masses opens up the most fundamental question of whether neutrinos are
Dirac or Majorana particles, which cannot be answered by neutrino oscillation
experiments.

Up to now, there is a consensus that the highest known sensitivity to small Majorana neutrino masses can be reached in experiments on the search of the $L$-violating neutrinoless double-$\beta$ decay process ($0\nu\beta\beta$)
\begin{equation}
	\label{eq:0nbb}
           ^AZ \to\, ^A(Z+2) + 2e^-\,,
\end{equation}
where $^AZ$ is a nucleus with atomic number $Z$ and mass number $A$. This
observable is proportional to the key parameter
\begin{equation}
	\label{eq:mbb}
	m_{\beta\beta} \equiv \sum_i  U^2_{ei}\, m_{\nu_i}\,,
\end{equation}
which is a coherent combination of the three neutrino masses.

There is an alternative to $0\nu\beta\beta$ by means of the mechanism of neutrinoless double electron capture ($0\nu\text{ECEC}$),
\begin{equation}
	\label{eq:0necec}
	^AZ + 2 e^- \to\, ^A(Z-2)^* \,.
\end{equation}

This is actually a mixing between two states of two different neutral atoms
differing in the total lepton number $L$ by two units, and the same baryonic
number $A$, and not a process conserving energy and momentum in general.
The daughter atom is in an excited state with two electron holes and its decay
provides the signal for (\ref{eq:0necec}). We study the implications of this
mixing between parent and daughter atom.

\section{The evolution Hamiltonian}
In the basis of the $\ket{^AZ\,}$ and $\ket{^A(Z-2)^*}$ states, which we'll refer to as 1 and 2, the dynamics of this two-state system of interest is governed by the Hamiltionian
\begin{equation}
	\dsH = \dsM - \frac{i}{2}\, \dsGamma
	= \left[ \begin{aligned}
			M_1 &\hspace{0.2cm} M_{21}^* \\
			M_{21} &\hspace{0.2cm} M_2
	\end{aligned} \right]
	- \frac{i}{2} \left[ \begin{aligned}
			0 &\hspace{0.2cm}0 \\
			0 &\hspace{0.2cm}\Gamma
	\end{aligned} \right] \,,
		\label{eq:H}
\end{equation}
with a Majorana $\Delta L = 2$ mass mixing $M_{21}$ from Eq.(\ref{eq:0necec}). 
The anti-Hermitian part of this Hamiltonian is due to the instability of $\ket{^A(Z-2)^*}$, 
which de-excites into $\ket{^A(Z-2)_\text{g.s.}}$, external to the two-body system in Eq.(\ref{eq:H}),
emitting its two-hole characteristic X-ray spectrum.

Besides being non-Hermitian, $\dsH$ is not a normal operator, i.e. $[\dsM,\,
	\dsGamma] \neq 0$. As a consequence, $\dsM$ and $\dsGamma$ are not compatible. The states of definite time evolution, eigenstates of $\dsH$, have complex eigenvalues and are given in non-degenereate perturbation theory \cite{galindo} by

\begin{align}
	\nonumber \ket{\lambda_L} &= \ket{1} + \alpha \ket{2}, \hspace{2cm} \lambda_L \equiv E_L -\frac{i}{2}\, \Gamma_L = M_1 + \abs{\alpha}^2 \left[ \Delta - \frac{i}{2}\, \Gamma \right] , \\
	\label{eq:eigen} \ket{\lambda_S} &= \ket{2} - \beta^* \ket{1}, \hspace{1.9cm} \lambda_S \equiv E_S -\frac{i}{2}\, \Gamma_S = M_2 -\frac{i}{2}\,\Gamma - \abs{\alpha}^2 \left[ \Delta - \frac{i}{2}\, \Gamma \right],
\end{align}
with $\Delta = M_1 - M_2$. As seen in Eq.(\ref{eq:eigen}), $\Gamma_{L,S}$ are \textbf{not} the eigenvalues
of the $\dsGamma$ matrix. The eigenstates are modified at first order in $M_ {21}$,
\begin{equation}
	\label{eq:alpha}
	\alpha = \frac{M_{21}}{\Delta + \frac{i}{2}\,\Gamma}\,,\hspace{2cm} 
	\beta = \frac{M_{21}}{\Delta - \frac{i}{2}\,\Gamma}\,,
\end{equation}
so the ``stationary'' states of the system don't have well-defined atomic properties: both the number of electrons and their atomic properties are a superposition of $Z$ and $Z-2$. Also, these states are \emph{not} orthogonal---their overlap is given by
\begin{equation}
	\braket{\lambda_S}{\lambda_L} = \alpha - \beta =  -i\frac{M_{21}\Gamma}{\Delta^2 + \frac{1}{4}\,\Gamma^2}\,,
\end{equation}
with its non-vanishing value due to the joint presence of the mass mixing $M_{21}$ and the decay width $\Gamma$. 
Notice that Im$(M_{21})$ originates a real overlap.

As seen in Eq.(\ref{eq:eigen}), the modifications in the corresponding eigenvalues appear at second order in $\abs{M_{21}}$ and they are equidistant with opposite sign. Since these corrections are small, from now on we will use the values
\begin{align}
	\nonumber E_L &\approx M_1\,, \hspace{2.375cm} E_S \approx M_2\,,\\
	\label{eq:GammaLS}  \Gamma_L &\approx \abs{\alpha}^2\, \Gamma\,,\hspace{2cm} \Gamma_S \approx \Gamma\,.
\end{align}
The only relevant correction at order $\abs{\alpha}^2$ is the one to $\Gamma_L$, since $\ket{1}$ was a stable state---even if it's small, the mixing produces a non-zero decay width.

This result shows that, at leading order, the Majorana mixing becomes observable through $\Gamma_L \propto \abs{\alpha}^2$. The value of $\alpha$ in Eq.(\ref{eq:alpha}) emphasizes the relevance of the condition $\Delta \sim \Gamma$, which produces a Resonant Enhancement \cite{17} of the effect of the $\Delta L=2$ mass mixing $M_{21}$.

\section{Natural time history for initial $^AZ$}
As seen in Eq.(\ref{eq:eigen}), the states $\ket{^AZ\,}$ and $\ket{^A(Z-2)^*}$ are not the stationary states of the system. For an initially prepared $\ket{^AZ\,}$, the time history is far from trivial and the appropriate language to describe the system short times after is that of \emph{Atom Oscillations} between $\ket{^AZ\,}$  and $\ket{^A(Z-2)^*}$ due to the interference of the amplitudes through $\ket{\lambda_S}$ and $\ket{\lambda_L}$ in the time evolution. 
The time-evolved $\ket{^AZ\,}$ gives raise to
the appearance probability
\begin{equation}
	\label{eq:t1}
	\abs{\braket{^A(Z-2)^*}{^AZ(t)}}^2 = \abs{\alpha}^2 \left\{ 1 + e^{-\Gamma t} - 2e^{-\frac{1}{2} \Gamma t}\cos(\Delta \cdot t) \right\}\,,
\end{equation}
with an oscillation angular frequency $\abs{\Delta}$. The characteristic oscillation time $\tau_\text{osc} = \abs{\Delta}^{-1}$ is the shortest time scale in this system. For $t \ll \tau_\text{osc}$, one has
\begin{equation}
	\abs{\braket{^A(Z-2)^*}{^AZ(t)}}^2 \approx \abs{M_{21}}^2\, t^2
\end{equation}
induced by the mass mixing. 

The next shortest characteristic time in this system is the decay time $\tau_S
= \Gamma^{-1}$, associated to the $\ket{\lambda_S}$ state. For $\tau_\text{osc}
\ll t \ll \tau_S$, the only change with respect to Eq.(\ref{eq:t1}) is that the
interference region disappears, and the two slits $\ket{\lambda_L}$ and
$\ket{\lambda_S}$ contribute incoherently,
\begin{equation}
	\abs{\braket{^A(Z-2)^*}{^AZ(t)}}^2 \approx \abs{\alpha}^2\,\left( 2-\Gamma t \right)\,.
\end{equation}

For $t\gg \tau_S$, the contribution of $\ket{\lambda_S}$ disappears and the appearance probability simply becomes
\begin{equation}
	\abs{\braket{^A(Z-2)^*}{^AZ(t)}}^2 = \abs{\alpha}^2\,.
\end{equation}
In other words, the initially prepared $\ket{^AZ\,}$ state evolves towards the stationary metastable state $\ket{\lambda_L}$,
\begin{equation}
	\label{eq:tl}
	\ket{^AZ(t)} \to e^{-i\lambda_L t}\ket{\lambda_L}\,,
\end{equation}
with the long lifetime $\tau_L = \Gamma_L^{-1}$ from Eq.(\ref{eq:GammaLS}). For
a realistic time resolution $\delta t$ in an actual experiment, this regime is
the interesting one, with the behavior in Eq.(\ref{eq:tl}). 
The different time scales involved in this problem are thus
\begin{equation}
	\tau_\text{osc} \ll \tau_S \ll \delta t \ll t \ll \tau_L \,, 
\end{equation}
where $t$ refers to the elapsed time since the production of $\ket{^AZ\,}$, either by nature or in the lab---given the smallness of the mixing, the metastability of the state (\ref{eq:tl}) is valid even for cosmological times. Therefore, for any time between the two scales $\tau_S$ and $\tau_L$, the populations of the three states involved are given by the probabilities\\
\begin{minipage}{0.75\textwidth}
\vspace{-0.8cm}\begin{align*}
	\mbox{}\\
	\hspace{3cm}\tau_S \ll t \ll \tau_L \hspace{0.5cm} \Longrightarrow \hspace{0.5cm}
	\left\{\begin{aligned}
		P_L(t) &\approx 1-\Gamma_L\, t\\
		P_S(t) &\approx 0\\
		P_\text{g.s.}(t) &\approx \abs{\alpha}^2\, \Gamma\, t
	\end{aligned} \right. \\
	\mbox{} 
\end{align*}
\end{minipage}
\hfill
\begin{minipage}{0.15\textwidth}
\vspace{-0.88cm}
\begin{subequations}
	\label{eqs:P}
	\begin{align}
		\mbox{} \label{eq:PL}\\
		\mbox{} \label{eq:PS}\\
		\mbox{} \label{eq:Pgs}
	\end{align}
\end{subequations}
\end{minipage} \\
where $P_\text{g.s.}(t)$ refers to the population of the ground state of the $^A(Z-2)$ atom after the decay of the unstable ``stationary'' state $\ket{\lambda_S}$, with rate $\Gamma$. No matter whether $t$ refers to laboratory or cosmological times, the linear approximation in $t$ is excellent.

\section{Observables}
With this spontaneous evolution of the system, an experiment beginning its measurements 
a time $t_0$ after the $^AZ$ was produced will probe the three-level system 
with relative populations $P_L \approx 100\%,\, P_S \approx 0,\, 
P_\text{g.s.}\approx \abs{\alpha}^2\Gamma\, t_0$. 
We discover different methods, involving the third state beyond the mixed states,
to be sensitive to the resonant Majorana mixing of atoms:
\begin{itemize}
	\item {\bf Spontaneous Emission from the metastable state to the daughter
		atom ground state.} The population in the upper level
		$\ket{\lambda_L}$, as shown in Eq.(\ref{eq:PL}), decreases with time as
		$P_L(\Delta t) \approx 1 - \Gamma_L\, \Delta t$, where $\Delta t = t -
		t_0$, due to the decay of the metastable ``stationary'' state
		$\ket{\lambda_L}$ to $\ket{^A(Z-2)_\text{g.s.}}$. This process is
		associated to the spontaneous emission of X-rays with a rate
		$\Gamma_L$, considered in the literature after the concept of Resonant
		Mixing was introduced in Ref.\cite{17}. For one mole of $^{152}$Gd, the
		X-ray emission rate would be of order $10^{-12}$~s$^{-1}\sim
		10^{-5}$~yr$^{-1}$. Its unique
		signature is the total energy of the two-hole X-ray radiation,
		displaced by $\Delta$ with respect to the
		characteristic $\ket{^A(Z-2)^*}\to \, \ket{^A(Z-2)_\text{g.s.}}$ X-ray
		spectrum, i.e. its energy release is the $Q$-value between the two
		atoms in their ground states.

	\item {\bf\boldmath Stimulated Emission from $\ket{\lambda_L}$ to
		$\ket{^A(Z-2)_\text{g.s.}}$.} The natural population inversion between
		the ground state and the metastable ``stationary'' state
		$\ket{\lambda_L}$ gives raise to the possibility of stimulating the
		decay $\ket{\lambda_L} \to \ket{^A(Z-2)_\text{g.s.}}$. The experimental
		signature of this process would be the emission of X-rays with total
		energy equal to the $Q$-value of the process, just like in the first
		observable of spontaneous emission.

		When compared with the spontaneous rate, one finds a gain factor
		\begin{equation}
		G = \hbar\, (\hbar c)^2\, \frac{\pi^2}{(\hbar \omega)^3}\, \frac{\dd N}{\dd{t}\dd{S}}\,
			\left[ \frac{\dd \omega}{\omega} \right]^{-1}\,,
			\label{eq:Gain}
		\end{equation}
		where $\dd N/\dd t\dd S$ is the luminosity $\cal L$ of the beam and
		$\omega$ the transition angular frequency.
		
		At XFEL, a sound simulation of the conditions of the machine \cite{altarelli} gives,
		for typical energies of tens of keV,
		the expected number of photons per pulse duration $dN/dt = 10^{10}$~fs$^{-1}$
		and the spectral width $\dd\omega/\omega = 1.12\times 10^{-3}$.
		Nanofocusing of this X-ray FELs has been contemplated \cite{gain};
		using a beam spot of the order of 100~nm would lead to a gain factor
		from (\ref{eq:Gain}) of $G\sim 100$. Applying this beam on a target of
		one mole of $^{152}$Gd would provide an X-ray emission rate of order
		$10^{-10}$~s$^{-1}\sim 10^{-3}$~yr$^{-1}$.

	\item {\bf Daughter Atom Population.} The presence of the daughter atom
		in the parent ores (see
		Eq.\ref{eq:Pgs}), can be probed e.g. by geochemical methods. For one mole of the
		nominally stable $^{152}$Gd isotope produced at the time of the Earth
		formation, one would predict an
		accumulated number of order $10^4$ $^{152}$Sm atoms.
		This observable could be of interest for cosmological times
		$t_0$ since, contrary to $\beta\beta$-decay, in the ECEC case there is
		no irreducible background from a $2\nu$ channel for a resonant atom
		mixing.

		In the presence of a light beam, the daughter atom population would absorb
		those characteristic frequencies corresponding to its energy levels, which would then
		de-excite emitting light of the same frequency.	In the case of the one mole
		$^{152}$Gd ore that we mentioned in the previous section, all $10^4$ Sm
		atoms could be easily excited to any of its $\sim 1$~eV levels using a standard pulsed laser of order
		$100$~fs pulse duration, with a mean power of $5$~W and a pulse rate of 
		$100$~MHz.

\end{itemize}

It is worth noting from the results of this section that the bosonic nature of
the atomic radiation is a property that can help in getting observable rates of
the Atom Majorana Mixing, including the stimulated X-ray emission from the
parent atom as well as the detection of the presence of the daughter atoms by
means of its characteristic absorption lines. The actual values correspond to
the specific case of $^{152}$Gd~$\to$~$^{152}$Sm, which is still off the Resonance 
Condition by at least a factor 30, implying a factor $10^3$ in the rates.

On this regard, there are many experimental searches 
[7-11]
of candidates
with better fulfillment of the resonance condition,
using the trap technique for precission measurement of atomic masses.

\section{Conclussions}
Neutrinoless double electron capture in atoms is a quantum mixing
mechanism between the neutral atoms $^AZ$ and $^A(Z-2)^*$ with two electron holes.
It becomes allowed for Majorana neutrino mediation responsible 
of this $\Delta L = 2$ transition.
This Majorana Mixing leads to the 
X-ray de-excitation of the $^A(Z-2)^*$ daughter atomic state which,
under the resonance condition, has no Standard Model background
from the two-neutrino decay.

The intense experimental activity looking for atomic candidates
satisfying the resonance condition by means of precise measurements of atomic masses, 
thanks to the trapping technique, has already led to
a few cases of remarkable enhancement effects and there is still room 
for additional adjustements of the resonance condition. With this
situation, it is important to understand the complete time evolution
of an atomic state since its inception and whether one can find,
form this information, different signals of the Majorana Mixing, including the
possible enhancement due to the bosonic nature of atomic transition radiation.
These points have been addressed in this work.

The effective Hamiltonian for the two mixed atomic states
leads to definite non-orthogonal states of mass and lifetime, each of them violating Global Lepton Number,
one being metastable with long lifetime and the other being short timelived.
For an initial atomic state there are time periods of Atom Oscillations, 
with frequency the mass difference, and the decay of the short lived state, which are not observable.
For observable times, the system of the two atoms has three relevant states for discussing transitions: 
one highly populated state with long lifetime, 
one empty state with short lifetime 
and the ground state of the daughter atom with a small population as a result of the past history. 
As a consequence, this is a case of natural population inversion suggesting the possibility 
of stimulated radiation transitions besides the natural spontaneous X-ray emission. 

The actual results obtained in this work demonstrate that the gain factors
which
could be obtained for double electron capture signals, by using the strategy of
stimulating the relevant  transitions,
are significant.
Taking into account the
ongoing searches for new atomic candidates with a better fulfillment of the resonance
condition, these processes could become realistic alternatives in the quest for the
Dirac/Majorana nature of neutrinos.

\ack
This research has been supported by MINECO Project 
FPA 2014-54459-P, Generalitat Valenciana Project GV 
PROMETEO II 2013-017 and  Severo  Ochoa  Excellence  
Centre  Project  SEV  2014-0398. AS acknowledges the 
MECD support through the FPU14/04678 grant.


\section*{References}
\begin{thebibliography}{9}
	\bibitem{1}
	Super-Kamiokande Collaboration (Fukuda Y et al.) 
	1998
	\emph{Phys. Rev. Lett.} {\bf 81} 1562

	\bibitem{2}
	SNO Collaboration (Ahmad Q R et al.)
	2002
	\emph{Phys. Rev. Lett.} {\bf 89} 011301
	
	\bibitem{galindo}
	Galindo A and Pascual P
	1990
	\emph{Quantum Mechanics}
	(Berlin: Springer)
	
	\bibitem{17}
	Bernabeu J, De Rujula A and Jarlskog C
	1983
	\emph{Nucl. Phys.} B {\bf 223} 15.
	
	\bibitem{altarelli}
	Schneidmiller E and Yurkov M, DESY note, private communication from Altarelli M
	
	\bibitem{gain}
	Yamauchi K, Yabashi M, Ohashi H, Koyama T and Ishikawa T
	2015
	\emph{J. Synchrotron Rad.} {\bf 22} 592,

	\bibitem{18}
	Eliseev S et al
	2011
	\emph{Phys. Rev. Lett} {\bf 106} 052504
	
	\bibitem{19}
	Kolhinen V S et al
	2010
	\emph{Phys. Lett.} B {\bf 684} 17
	
	\bibitem{21}
	Barabash A S et al
	2008
	\emph{Nucl. Phys.} A {\bf 807} 269
	
	\bibitem{24}
	Goncharov M et al
	2011
	\emph{Phys. Rev.} C {\bf 84} 028501
	
	\bibitem{25}
	Droese C et al
	2012
	\emph{Nucl. Phys.} A {\bf 875} 1


\end{thebibliography}
\end{document}